\documentclass[11pt,preprint]{aastex}

\begin{document}
\title{Taking ``The Road Not Taken'':\\ On the Benefits of
Diversifying Your Academic Portfolio\footnote{Write-up of a
banquet lecture at the conference on ``The First Galaxies, Quasars,
and Gamma-Ray Bursts,'' Penn State University (June 2010).}}

\author{Abraham Loeb\\Institute for Theory \& Computation\\
Harvard University\\60 Garden St., Cambridge, MA 02138}

\begin{abstract}

It is common practice among young astrophysicists these days to invest
research time conservatively in {\it mainstream} ideas that have
already been explored extensively in the literature.  This tendency is
driven by peer pressure and job market prospects, and is occasionally
encouraged by senior researchers.  Although the same phenomenon
existed in past decades, it is alarmingly more prevalent today because
a growing fraction of observational and theoretical projects are
pursued in large groups with rigid research agendas. In addition, the
emergence of a ``standard model'' in cosmology (albeit with unknown
dark components) offers secure ``bonds'' for a safe investment of
research time.  In this short essay, which summarizes a banquet
lecture at a recent conference, I give examples for both safe and
risky topics in astrophysics (which I split into categories of
``bonds,'' ``stocks,'' and ``venture capital''), and argue that young
researchers should always allocate a small fraction of their academic
portfolio to innovative projects with risky but potentially highly
profitable returns. In parallel, selection and promotion committees
must find new strategies for rewarding candidates with creative
thinking.

\end{abstract}

\bigskip
\bigskip
\bigskip

{\it ~\\``Two roads diverged in a wood, and I -- \\
I took the one less traveled by, \\
And that has made all the difference.''}

\noindent
{\it from the poem ``The Road Not Taken'' by Robert Frost.}


\section{Introduction}

It is impossible to forecast the scientific truths that will be
unraveled by future generations of astrophysicists looking at the sky,
since the Universe is often more subtle than our imagination. Attempts
to establish a dogmatic view about the sky have often failed; notable
examples include the notion that the Sun moves around the Earth, that
the observable Universe has existed forever, or that there is little
to be found in the X-ray sky. The history of our profession teaches us
modesty. We cannot pretend that we have the final answers, especially
at a time when most of the content of our Universe is attributed to
unknown entities (dark matter and dark energy), and most of the
comoving volume of the observable Universe has not been mapped
yet\footnote{A. Loeb, ``When Did the First Stars and Galaxies Form?'',
Princeton University Press (2010).} (see Fig. 1).  Under such
circumstances, we should allow ourselves to think from time to time
outside the (simulation) box. It is always prudent to allocate some
limited resources to innovative ideas beyond any dogmatic
``mainstream,'' because even if only one out of a million such ideas
bears fruit, it could transform our view of reality and justify the
entire effort. This lesson is surprisingly unpopular in the current
culture of funding agencies like NSF or NASA, which promote research
with predictable and safe goals.

\begin{figure*}[hptb]
\centerline{\includegraphics[scale=0.45,angle=-90]{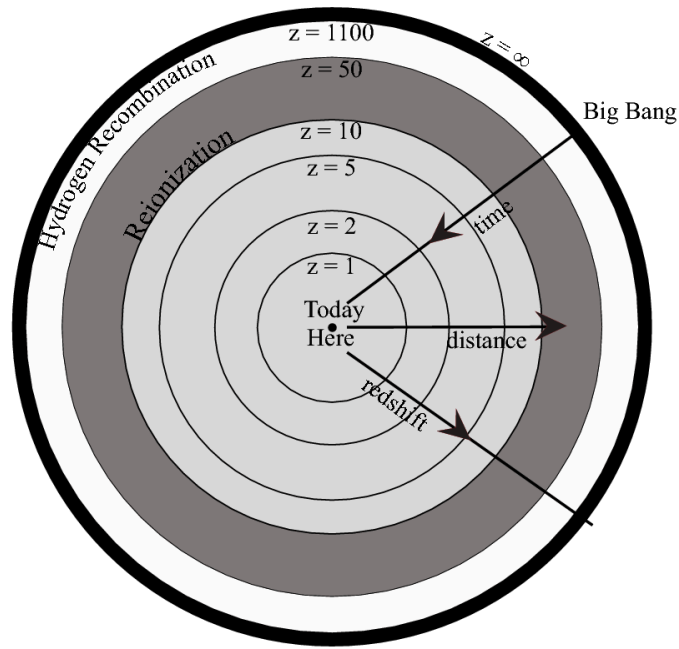}}
\caption{The distribution of matter through 99.9\% of the comoving
volume of the observable Universe (at redshifts $z>0.3$) has not
been mapped as of yet. We should therefore avoid being dogmatic for
now.}
\label{Fig1}
\end{figure*}

Personally, I enjoy doing science for the rare thrill of discovering
something new about the Universe that has never been understood
before.  By its nature, creative work is unexpected and provides less
security than, say, the predictable task of a ``mainstream'' engineer
who uses existing knowledge to construct a bridge. The required
day-to-day effort is also no fun: in order to find one new idea that
works, I usually search through numerous other ideas that either fail
or have already been examined in the literature before. As a result,
the fraction of my time dedicated to innovation is far greater than
the fraction of my papers which are innovative.  Although experience
helps to cull out bad ideas without spending too much time on them,
experience is also a double-edged sword since prejudice could
compromise a rare opportunity for discovery.

New discoveries are naturally made in unexplored territories. Young
people are more capable of exploring the ``roads not taken'' because
they lack an unwarranted baggage of prejudice (or adopt a flat
Bayesian prior) on the likelihood of discovery along these roads. The
window of opportunity in a scientist's career is often short: after
tenure, most senior researchers get distracted by administrative and
fund-raising concerns, and prefer to maintain a conservative profile
that promotes old ideas within their discipline.

Clearly, failure and waste of time are a common outcome of risky
projects, just as the majority of {\it venture capital} investments
lose money (but have the attractive feature of being more profitable
than anything else if successful).  The fear of losses is sure to keep
most researchers away from risky projects, which will attract only
those few who are willing to face the strong headwind.  Risky projects
are accompanied by loneliness. Even after an unrecognized truth is
discovered, there is often persistent silence and lack of attention
from the rest of the community for a while. This situation contrasts
with the nurturing feedback that accompanies a project on a variation
of an existing theme already accepted by a large community of
colleagues who work on the same topic. Martin Schwarzschild told me
that in the 1950s most astronomers were working on binary stars, and
conferences were filled with talks that sounded just like each other
on this popular topic. According to him, the sociology of astrophysics
has not changed since then -- only the title of the current ``topic of
the day''.

Indeed, conformism is not new. But the astrophysics community is
bigger now than it used to be, with stronger social pressure and more
competition in the job market. These forces exaggerate the herd
mentality to an extent that suggests a need for policy change by our
funding agencies. A growing fraction of observational and theoretical
projects are done in large groups with rigid research agendas and
tight schedules.  While the mode of large groups has dominated
experimental particle physics for decades, it has become popular in
observational and theoretical astrophysics only
recently.\footnote{Here I define a ``large group'' to be one that
produces research papers in which the list of authors is longer than
the abstract.}  In addition, the emergence of a ``standard model'' in
cosmology offers secure research ``bonds'' in which young cosmologists
can invest their time with minimal risk.

The purpose of this write-up is to encourage young researchers to
resist this alarming trend and pursue innovative research, and to
encourage senior members sitting on selection, promotion, and grant
awarding committees to find better strategies for rewarding
creativity.  A change in attitude is crucial for the future health of
our field.

\section{Risk management and historical trends}

Just as in monetary investments, the level of risk that a researcher
chooses to adopt is a matter of both personal inclination and social
factors.  Figure 2 shows present-day examples of topics that belong to
categories of low-risk (``bonds''), medium risk (``stocks''), or
high-risk (``venture capital'') investments of research time.

\begin{figure*}[hptb]
\centerline{\includegraphics[scale=0.45,angle=-90]{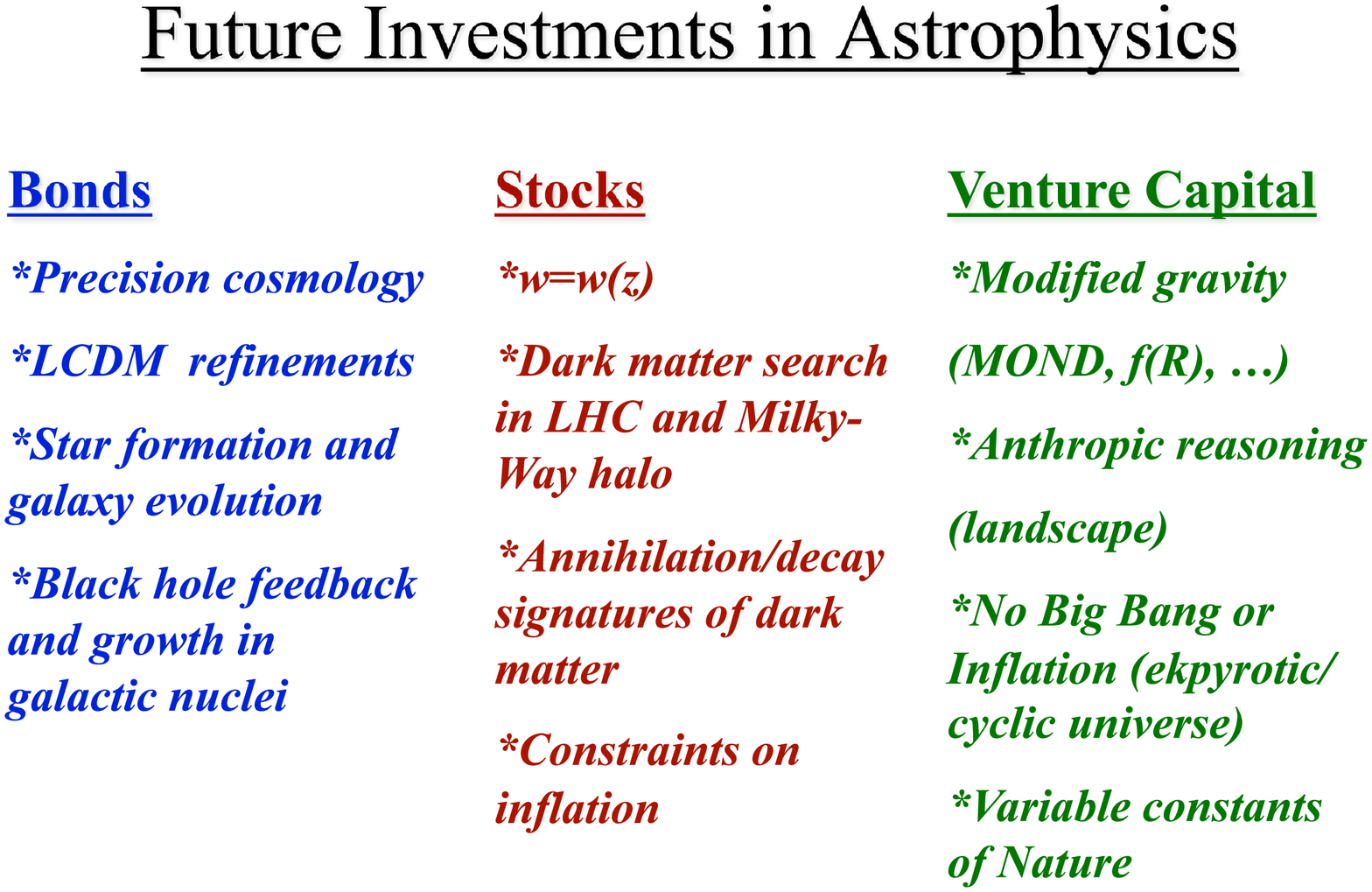}}
\caption{Categories of risk management in astrophysics research.  Each
topic includes components with different shades of risk, some of which
deviate from the listed category of the topic's centroid.}
\label{Fig2}
\end{figure*}

Timing is crucial for making a profit. When should a young researcher
be more inclined to invest in a bond, stock, or venture capital?
Common sense suggests that the answer depends on the state of the
field.  An illustrative example is provided by the choice between dark
matter and modified gravity. Once experiments push the upper
limit on the cross-section of Weakly-Interacting Massive Particles
(WIMPs) down well below the expectation of most reasonable models,
alternative gravity models might appear more appealing.

When I was a postdoc two decades ago, there was no standard model in
cosmology, and it was more socially acceptable for young cosmologists
to invest in ``venture capital'' ideas.  Today, young cosmologists are
mainly investing in bonds with the premise that they offer greater
security. We should keep in mind that historically there were
trajectories of ideas that started as ``venture capital,'' turned into
``stocks,'' and eventually matured into ``bonds.'' Examples include:
\begin{itemize}
\item {\bf The Big Bang} 
\item {\bf The Cosmic Microwave Background (CMB)} 
\item {\bf Dark Matter} 
\item {\bf Inflation}
\end{itemize}
But the risks should also be recognized. In other cases, ideas that
started as venture capital turned into ``junk bonds.'' Examples
include:
\begin{itemize}
\item {\bf Steady-state cosmology} 
\item {\bf Topological defects} 
\item {\bf Dark matter is baryonic} 
\item {\bf X-ray background from a hot intergalactic medium} 
\item {\bf Quintessence}
\end{itemize}

The possible trajectories of innovative ideas in astrophysics are
summarized in Figure 3.

\begin{figure*}[hptb]
\centerline{\includegraphics[scale=0.4,angle=-90]{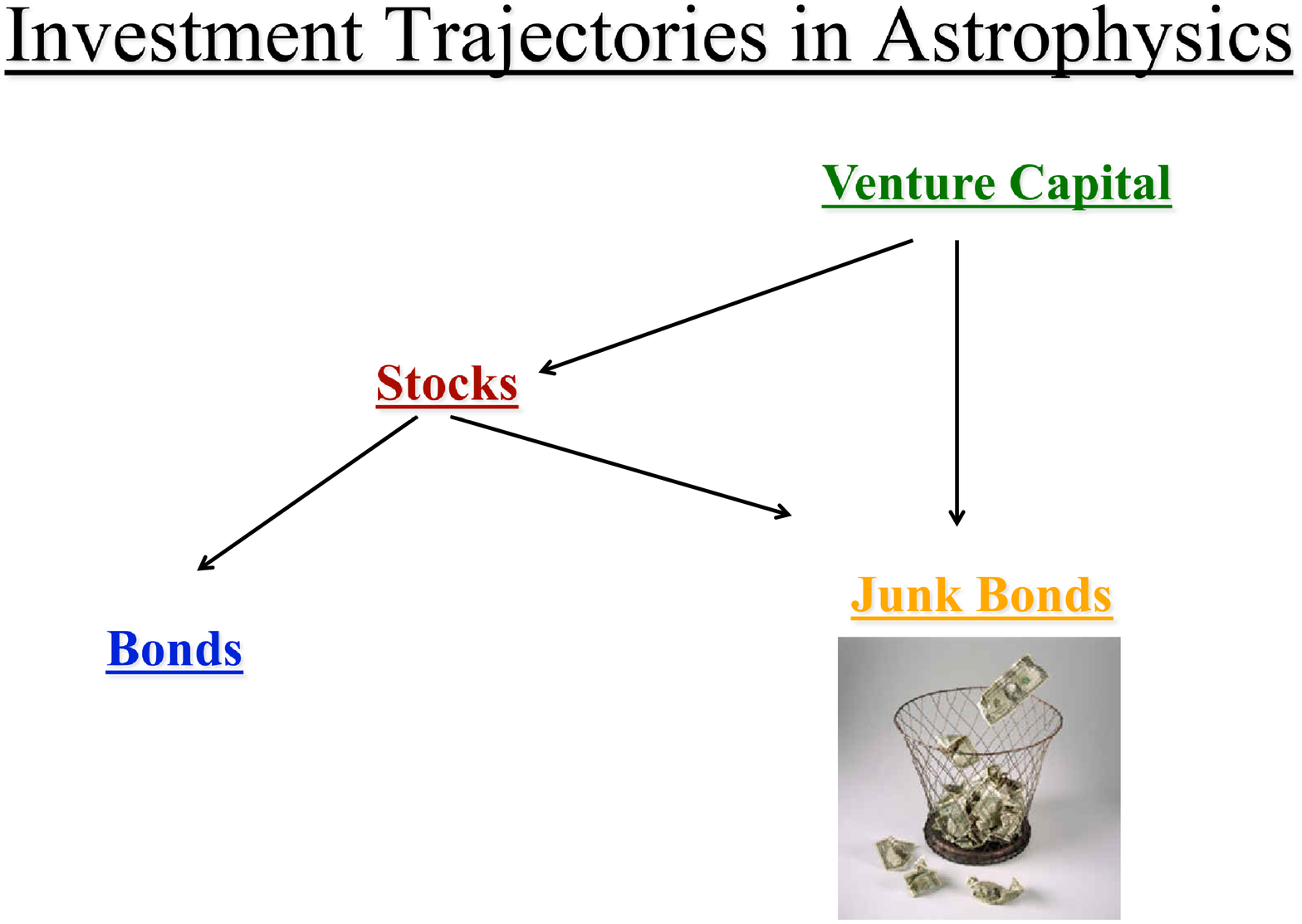}}
\caption{Historical trajectories of ideas in astrophysics.}
\label{Fig3}
\end{figure*}

\section{Examples of Future Frontiers in Astrophysics}

To illustrate how future frontiers in astrophysics may differ from
traditional ones, let me list a few concrete examples.  Given that
science often brings surprises, my list is not meant to represent a
forecast but rather a sample of possibilities. I am fully aware of
Rabbi Yochanan's wise disclaimer reported in the Talmud: ``following
the destruction of the temple, prophecy became the trademark of fools
and babies.''

\noindent
\underline{1. In the long-term future (centuries from now):}
\begin{itemize}
\item {\bf Biology away from the Earth:} life may be explored not only
directly in our solar system (fish in Europa, enceladus), but also
remotely in habitable extra solar planets.  If mature civilizations
exist, their signals may be detected. Astrophysicists may understand
better the dark matter and dark energy based on the science done by
these civilizations over the past billions of years (although this
might feel like cheating in an exam). The origin of life might be
explored in parallel in the laboratory.
\item {\bf Laboratory cosmology:} human-made experiments might
attempt to study the inflaton or perturb the dark energy.
\end{itemize}

\noindent
\underline{2. In the short-term future (decades from now):}
\begin{itemize}
\item {\bf Dark matter experiments} might produce
(LHC\footnote{http://lhc.web.cern.ch/lhc/}) or directly detect (from
the Milky Way halo) the dark matter.
\item {\bf Gravitational waves astrophysics:} Advanced
LIGO\footnote{http://www.ligo.caltech.edu/} will be used to study
neutron star-neutron star or neutron star-black hole mergers as the
origin of short gamma-ray bursts.
LISA\footnote{http://lisa.nasa.gov/} will map the growth history of
supermassive black holes out to the first galaxies at $z\sim 20$, and
test general relativity.
\end{itemize}

\noindent
\underline{3. In the immediate future (years from now):}

\begin{itemize}
\item {\bf Imaging black holes} will be achieved with Very Large
Baseline Interferometry (VLBI) at sub-millimeter
wavelengths\footnote{http://arxiv.org/abs/0906.3899}.
\item {\bf The 21-cm line of hydrogen in the 21st century} will be
used to map the gravitational growth of perturbations throughout most
of the observable volume of the Universe
(LOFAR\footnote{http://www.lofar.org/} and
MWA\footnote{http://www.MWAtelescope.org/}).
\item {\bf JWST\footnote{http://www.jwst.nasa.gov/} and the next
generation of large telescopes
(EELT\footnote{http://www.eso.org/sci/facilities/eelt/},GMT\footnote{http://www.gmto.org/},
and TMT\footnote{http://www.tmt.org/})} will identify fainter and
smaller sources of light than ever probed before.
\item {\bf Transient surveys (LSST\footnote{http://www.lsst.org/lsst}
Pan-STARRS\footnote{http://pan-starrs.ifa.hawaii.edu/public/} and
PTF\footnote{http://www.astro.caltech.edu/ptf/})} will discover new
types of explosions and variable sources.
\item {\bf Planet searches (e.g., with
Kepler\footnote{http://www.nasa.gov/mission$_{-}$pages/kepler/main/index.html})}
will discover habitable planets around nearby stars.
\end{itemize}

\section{Recommendations for the investment strategy of young researchers}

The most common investment strategy of research time by young postdocs
in astrophysics these days is:
\begin{itemize}
\item 80\% in bonds 
\item 15\% in stocks
\item 5\% in venture capital.
\end{itemize}
My recommended strategy is:
\begin{itemize}
\item 50\% in bonds
\item 30\% in stocks
\item 20\% in venture capital.
\end{itemize}
My other general suggestions are:
\begin{itemize}
\item {\bf Avoid ideas that are second-order speculations} (since the
probability of both ideas being right is the product of the small
probabilities that each of them is right). Examples include
non-Gaussianity from cosmic strings, formation of mini-black holes as
the dark matter at the end of inflation, proposing a process that
tuned the dark energy to have roughly the mean cosmic density today
and to evolve roughly on the Hubble time today, and using Modified
Newtonian Dynamics (MOND) $+$ neutrino dark matter to explain the CMB
anisotropies.
\item {\bf Progress is not linear in time.} If you uncover a new
unexpected finding at the end of your research project, study it
carefully; if it turns out to be important, write your main research
paper about it even though it was unplanned and you invested much more
time in the original project.  Notable examples include the unplanned
discovery of the CMB in background noise tests of a radio horn
antenna, or the accidental discovery of giant arcs associated with
gravitational lensing in X-ray clusters.
\item {\bf Leave spare time} for the unexpected discovery that 
might change your research plans altogether (once per decade).
\end{itemize}
Selection and promotion committees as well as grant awarding agencies
must find new ways to reward creative thinking. We will all benefit
richly from the implementation of a new strategy. And for those who
take ``the road not taken,'' keep your spirits up. {\it Is there
any point to doing science other than taking that road?}

\bigskip
\bigskip
\bigskip

\acknowledgements I thank the organizers of the Penn State conference
for assigning me the challenge of a banquet lecture early in my
career. Special thanks go to I. Liviatan, J. Pritchard, and G. White
for their helpful comments on the manuscript.

\end{document}